\documentclass[epj]{svjour}
\usepackage{epsfig,amsmath,amssymb}
\def\Jtd{J_{2,{\rm dim}}}

\begin{document}

\title{Jordan-Wigner approach to
the frustrated spin one-half $XXZ$ chain}

\author{T. Verkholyak\inst{1}
\and A. Honecker\inst{2}
\and W. Brenig\inst{2}
%
}                     
%
%
%
\institute{
Institute for Condensed Matter Physics,
1 Svientsitskii Str., L'viv-11, 79011, Ukraine
\and
Technische Universit\"at Braunschweig, Institut f\"ur Theoretische
Physik, Mendelssohnstrasse 3, 38106 Braunschweig, Germany
}
%
%
\date{November 16, 2005; revised January 6, 2006}
\abstract{
The Jordan-Wigner transformation is applied to study the ground
state properties and dimerization transition in the $J_1-J_2$
$XXZ$ chain. We consider different solutions of the mean-field
approximation for the transformed Hamiltonian. Ground state
energy and the static structure factor are compared with complementary
exact diagonalization and good agreement is found
near the limit of the Majumdar-Ghosh model.
Furthermore, the ground state phase diagram is discussed
within the mean-field theory. In particular, we show that an
incommensurate ground state is absent for large $J_2$ in a fully
self-consistent mean-field analysis.
}

\maketitle
\section{Introduction}
Low-dimensional quantum spin systems with competing interaction
are a field of diverse theoretical studies \cite{rice2001}.
Recent progress in material synthesis has brought up  several
transition metal oxides realizing such spin systems.
In particular, the $s=1/2$ spin chain
with frustrated next-nearest neighbor interaction
has been found to model the quasi one-dimensional copper
oxide material CuGeO$_3$ \cite{lemmens2003,hase2004}.
The $J_1-J_2$ Heisenberg chain is well known to display a quantum phase
transition from a gapless, translationally invariant ground state with
algebraic spin-correlations to a dimer state \cite{haldane1982} at
$J_2\approx 0.24 J_1$ \cite{white1996,eggert1996}.
At the Majumdar-Ghosh point \cite{MaGo}, i.e.\ at $J_2=J_1/2$,
the ground state is a doubly degenerate dimer product of
singlet pairs on neighboring sites. In the vicinity of the
Majumdar-Ghosh point there is a transition \cite{Nomura} to
an incommensurate phase for larger $J_2$ \cite{white1996,bursill1995}.
Low-energy field theories \cite{white1996,NGE98} are not suitable
for describing this transition since it occurs at a very short
correlation length of about one lattice spacing. Also semiclassical
approaches \cite{AS95,kolezhuk} are not applicable
to this commensurate-incommensurate transition in the
$J_1-J_2$ chain with $s=1/2$, since they fail
to reproduce e.g.\ the solitonic nature of the fundamental
excitations \cite{ShaSu}.

The Jordan-Wigner transformation is another technique
which allows for an approximate analytic
analysis of low-dimensional quantum spin systems
(see e.g.\ Ref.~\cite{derzhko2001} for a recent review).
It has also been applied to the $J_1-J_2$ chain \cite{brenig1997,sun2002},
however only for a restricted set of mean-field configurations.
While Ref.~\cite{sun2002} reported incommensurability for
larger $J_2$, this work did not consider dimerized states
and thus fails to reproduce the gap which is known to be
relevant in this region (see e.g.\ \cite{white1996}).
This motivates us to perform
a systematic study of the spin-$1/2$ $J_1-J_2$ chain using
the Jordan-Wigner transformation and considering all
possible mean-field configurations. Since the Jordan-Wigner
transformation is known to be exact for the nearest-neighbor
$XY$ chain \cite{takahashi}, we will further consider the
case of a general $XXZ$ anisotropy. Note that the effects
of the anisotropy in the spin-1/2 $J_1-J_2$ chain
have been studied systematically either numerically
\cite{nomura1993,somma2001} or using field theoretical
approaches \cite{haldane1982,j1j2xy,ZFN04}, but not in
previous Jordan-Wigner treatments \cite{brenig1997,sun2002}.

This paper is organized as follows: In section \ref{secMF}
we perform a mean-field analysis of the Jordan-Wigner
transformed \cite{jw,fradkin}
Hamiltonian. Remarkably, the Majumdar-Ghosh state
\cite{MaGo,caspers} is recovered exactly by our approach.
Section \ref{secRes} presents results obtained from a
numerical solution of the mean-field equations and a
comparison with complementary exact diagonalization data.
Finally, section \ref{secConc} summarizes our results
concerning in particular the applicability of the
Jordan-Wigner approach to the different parameter regimes.

\section{Mean-field analysis} 
\label{secMF}

In the following
we consider the frustrated spin-$1/2$ $XXZ$ chain of $L$ spins
($L\to\infty$):
\begin{eqnarray}
H=\sum_{l=1}^L J_1(s_l^xs_{l+1}^x+s_l^ys_{l+1}^y + \Delta s_l^zs_{l+1}^z)
\nonumber\\
+J_2(s_l^xs_{l+2}^x+s_l^ys_{l+2}^y + \Delta s_l^zs_{l+2}^z).
\label{ham}
\end{eqnarray}
Here $J_1$ ($J_2$) $>0$ is the nearest (next-nearest) neighbor coupling,
$\Delta$ is the exchange anisotropy
and $s_l^\alpha$ are spin-$1/2$  operators.
By the Jordan-Wigner transformation \cite{jw,fradkin} the spin
Hamiltonian (\ref{ham}) is mapped onto a model of interacting
spinless fermions $c_l^{(+)}$:
\begin{eqnarray}
H&=& \sum_{l=1}^L \bigg(\frac{J_1}{2}c_l^+c_{l+1}
+\frac{J_2}{2}c_l^+c_{l+2}
-J_2 c_l^+c_{l+1}^+c_{l+1}c_{l+2}
\nonumber\\
&+&h.c.\bigg)
+\Delta
J_1\left(c_l^+c_l-\frac{1}{2}\right)
\left(c_{l+1}^+c_{l+1}{-}\frac{1}{2}\right)
\nonumber\\
&+&\Delta J_2 \left(c_l^+c_l-\frac{1}{2}\right)
\left(c_{l+2}^+c_{l+2}{-}\frac{1}{2}\right).
\label{jwham}
\end{eqnarray}
The first term corresponds to the $XY$ nearest-neighbor interaction
and can be treated rigorously \cite{lsm}.
The remaining terms are the $z$ component of the nearest
neighbor and the complete next-nearest neighbor exchange.
They represent four-fermion interactions which render the
model non-integrable. We will therefore
resort to a mean-field approximation preserving
all pair correlations of type $\langle c_n^+ c_m\rangle$
in the factorization of the interaction terms:
\begin{eqnarray}
g_l&=&\langle s_l^z\rangle=\langle c_l^+c_l\rangle-\frac{1}{2},
\nonumber\\
A_l&=&\langle c_l^+c_{l+1}\rangle=\langle c_{l+1}^+ c_l\rangle,
\nonumber\\
D_l&=&\langle c_l^+c_{l+2}\rangle=\langle c_{l+2}^+ c_l\rangle.
\label{contractions}
\end{eqnarray}
These contractions are related to single-site, nearest neighbor,
and next-nearest neighbor spin-spin correlation functions.
The self-consistent determination of these contractions follows
Ref.~\cite{brenig1997}, where however only the case of $g_l=D_l=0$
had been considered. We find the following phases:
\begin{itemize}
\item[i)] paramagnetic ({\em homogeneous}) \cite{bul1962}:
$A_l=A=-\frac{1}{\pi}$, $g_l=D_l=0$
\item[ii)] uniform antiferromagnetic ({\em AFM+uni.}):
   $g_l=(-1)^l g$, $A_l=A$, and $D_l=D$
\item[iii)] staggered antiferromagnetic ({\em AFM+stag.}):
   in contrast to ii) $D_l$ is staggered, i.e. $D_l=(-1)^l D$
\item[iv)] alternating nearest neighbor hopping ({\em dimer}):
   $A_l=A+(-1)^l\delta$.
\end{itemize}
The self-consistency equations for phase iv) are discussed below.
Here, we should only note that in all cases we find no
uniform contribution to the next-nearest neighbor correlation
$D_l$. A staggered next-nearest neighbor contraction is induced by the
antiferromagnetic order and tends to zero when $g=0$.
Therefore, the current mean-field approach is not suited to
treat the limit of two weakly coupled chains ($J_1\to 0$)
where the correlation between the next-nearest neighbor spins becomes strongest.

For case iv), and after Fourier transformation
the mean-field Hamiltonian reads
\begin{eqnarray}
H=\sum_{-\pi<k\leq\pi}
e_k \, c_k^+c_k + \frac{if_k}{2} (c_{k\pm\pi}^+c_k-c_k^+c_{k\pm\pi}) \, ,
\label{approx}
\end{eqnarray}
where
$e_k=(J_1-2A(\Delta J_1-2J_2))\cos(k)$,
$f_k=2\delta(\Delta J_1+2J_2)\sin(k)$.
The Hamiltonian can be diagonalized by the unitary transformation
\begin{eqnarray}
c_k &=& \frac{1+i}{\sqrt{2}} (\cos(\alpha_k^f/2) \, \eta_k
- \sin(\alpha_k^f/2) \, \eta_{k\pm\pi}) \, ,
\nonumber\\
c_{k\pm\pi} &=& \frac{1-i}{\sqrt{2}} (\sin(\alpha_k^f/2) \, \eta_k
+ \cos(\alpha_k^f/2) \, \eta_{k\pm\pi}) \, ,
\nonumber
\label{transform}
\end{eqnarray}
where $\cos\alpha_k^f=e_k/\sqrt{e_k^2{+}f_k^2}$.
This leads to the free fermion model
\begin{eqnarray}
&&H=\sum_{-\frac{\pi}{2}<k\leq\frac{\pi}{2}}
\lambda_k \left(\eta_k^+\eta_k
-\eta_{k\pm\pi}^+\eta_{k\pm\pi}\right) \, ,
\label{diagham}
\end{eqnarray}
 with the spectrum:
$\lambda_k=\sqrt{e_k^2+f_k^2}$.
For the selfconsistent equations for $A$, $\delta$ we find:
\begin{eqnarray}
&&A=-\frac{1}{2\pi}\int_{-\frac{\pi}{2}}^{\frac{\pi}{2}}
dk \, \cos(k)\frac{e_k}{\sqrt{e_k^2+f_k^2}} \, ,
\nonumber\\
&&\delta=\frac{1}{2\pi}\int_{-\frac{\pi}{2}}^{\frac{\pi}{2}}
dk \, \sin(k)\frac{f_k}{\sqrt{e_k^2+f_k^2}}\, .
\label{eqs}
\end{eqnarray}
The mean-field approximation to the ground-state energy is obtained
from the ground state expectation value of the fermion Hamiltonian
(\ref{jwham}), including contractions only up to quadratic order.
For the dimer phase iv) the ground state energy per site is
\begin{eqnarray}
e=J_1A-(\Delta J_1-2J_2)A^2-(\Delta J_1+2J_2)\delta^2 \, .
\label{energy}
\end{eqnarray}
We note that the dimer order parameter \cite{white1996}
$d{=}\langle\vec{S}_{2i{-}1}\vec{S}_{2i}\rangle -
\langle\vec{S}_{2i}\vec{S}_{2i+1}\rangle$
corresponds to $4(A-1)\delta$ within our treatment.

So far, we have assumed that the mean fields are real. However,
it is known that an easy-plane anisotropy $\Delta<1$
induces a chiral phase for large $J_2$ \cite{NGE98,HKK01}
(note that the existence of such a chiral phase had
been verified numerically in a different ladder model \cite{Nishiyama}
prior to \cite{HKK01}).
The chiral order parameter
$\kappa_l=s_l^x s_{l+1}^y-s_l^y s_{l+1}^x$ corresponds
to the imaginary part of the fermion nearest-neighbor contraction
$\Im{\rm m} A_{l}$.
Therefore, the elementary contraction should be allowed to
be complex. In this case the ground state energy reads:
\begin{eqnarray}
e &= &J_1A_R-(\Delta J_1-2J_2)A_R^2-(\Delta J_1+2J_2)\delta_R^2
\nonumber\\
&&-(\Delta J_1+2J_2)A_I^2-(\Delta J_1-2J_2)\delta_I^2 \, .
\label{en-chiral}
\end{eqnarray}
Here $A_{R(I)}$ is the real (imaginary) part of $A$, and
$\delta_{R(I)}$ is the real (imaginary) part of $\delta$.
It is evident that for large $J_2/J_1$ a non-zero
$A_I$ lowers the ground state energy in contrast to $\delta_I$.
Hence, we have considered mean-field solutions with
non-zero $A_R$, $A_I$, and $\delta_R$. However, numerical
solution of the mean-field equations did not yield
any complex solution. 

The Majumdar-Ghosh point \cite{MaGo,caspers} $J_2={J_1}/{2}$
permits to check the consistency of the mean-field approximation.
At this point the ground state is doubly degenerate and consists
of singlet pairs on neighboring sites
$\prod_{l=1}^{N/2}[2l\mp1,2l]$, where $[2l\mp 1,2l]$ denotes one
singlet. This can be related to the fermion
language by recalling that $S^z=+(-)1/2$ on site $l$ corresponds
to a filled (empty) site $l$. Therefore $(c_{2l\mp 1}^+
-c_{2l}^+)|0\rangle$ creates a singlet bond on neighboring sites
$2l\mp 1$, $2l$, and the Majumdar-Ghosh states can be
represented \cite{buettnerUn} as
$\prod_{l=1}^{N/2} (c_{2l\mp 1}^+-c_{2l}^+) |0\rangle$.
We will now argue that this exact state is also obtained
from the mean-field solution. For $J_2 = J_1/2$
configuration iv) and Eqn.~(\ref{eqs}) yield
$A=-1/4$, $\delta=\pm 1/4$. Inserting this solution into the mean-field
Hamiltonian one gets
\begin{eqnarray}
H=\sum_{l=1}^{N/2}\frac{J_1(1+\Delta)}{2}
(\eta_l^{(t)+}\eta_l^{(t)}-\eta_l^{(s)+}\eta_l^{(s)})
+\frac{NJ_1\Delta}{8},
\end{eqnarray}
where $\eta_l^{(t)}=\frac{1}{\sqrt{2}}(c_{2l\mp1}+c_{2l})$,
$\eta_l^{(s)}=\frac{1}{\sqrt{2}}(c_{2l\mp1}-c_{2l})$ for
$\delta=\pm\frac{1}{4}$.
$\eta_l^{(s)+}$ and $\eta_l^{(s)}$ ($\eta_l^{(t)+}$ and $\eta_l^{(t)}$)
create and annihilate a singlet (triplet) on nearest neighbors.
This shows that the mean-field solution is
the Majumdar-Ghosh state in fermionic representation
$\prod_{l=1}^{N/2} \eta_l^{(s)+}|0\rangle$
with the ground state energy per spin $e=-J_1(2+\Delta)/8$.

\section{Results}
\label{secRes}

\begin{figure}[t!]
\centering
\epsfig{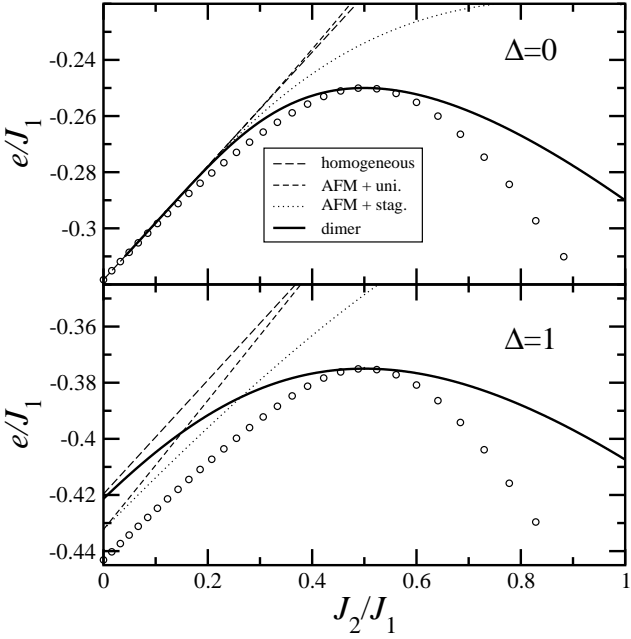}
\caption[]
{Ground-state energy per site as a function of $J_2/J_1$
for the $J_1-J_2$ $XY$ chain (upper panel) and the Heisenberg
chain (lower panel):
the circles are exact diagonalization data extrapolated to
the infinite system;
lines correspond to the different mean-field solutions for the ground state.
}
\label{gs-energy}
\end{figure}

Fig.\ \ref{gs-energy} shows results of a numerical solution of
the self-consistency equations for configurations i)-iv). It
can be seen that all solutions exist for positive $J_2$. To
check the accuracy of these results, we have performed exact
diagonalization of the Hamiltonian (\ref{ham}) for $L \le 32$
sites (since this type of computations has a long
tradition \cite{tonegawa1987}, similar results can also be found
in the literature, in particular for $\Delta  = 1$). We have
performed a finite-size extrapolation using the
Vanden-Broeck-Schwartz-algorithm \cite{VBS79} with
$\alpha_{\rm VBS} = -1$ and, for most cases, $L = 12$, $16$,
\ldots $32$. The
extrapolated results for the ground-state energy per site
$e$ are shown by the circles in Fig.\ \ref{gs-energy}. Errors
are estimated to be smaller than the size of the symbols.
For $J_2 = 0$, our extrapolated values agree to better
than $10^{-6} J_1$ with
the  exact answer \cite{takahashi}
$e/J_1 = -1/\pi \approx -0.318310$  and
$1/4 - \ln(2) \approx -0.443147$ for
$\Delta = 0$ and $1$, respectively.

Apart from the exact ground state at
the Majumdar-Ghosh point we may contrast the mean-field
solutions against other known results. We start
from the $XY$ nearest-neighbor model ($\Delta=0$, $J_2=0$)
where the Jordan-Wigner transformation yields the exact
answer \cite{takahashi}.
For small $J_2/J_1$ the results of all
approximations are very close (upper panel of Fig.~\ref{gs-energy})
and all order parameters grow very slowly.
In fact, a detailed analysis of (\ref{eqs}) for $A$, $\delta$
shows that $J_2>\Jtd=-\Delta J_1/2$
is the condition of a non-zero solution for $\delta$.
The approximate solution for small $J_2-\Jtd$ above this point gives
\begin{eqnarray}
\delta\approx\frac{a}{e(J_2-\Jtd)}
\exp\left(-\frac{\pi a}{4(J_2-\Jtd)}\right),
\label{asymptotic}
\end{eqnarray}
where $a=J_1-\frac{4}{\pi}(J_2+\Jtd)$.
The excitation gap appears at $k=\pi/2$ and increases as
$4(J_2-\Jtd)\delta$. Note that close to the critical point ${J_2}_c$
bosonization yields a very similar form \cite{haldane1982,white1996}
$\delta\sim \exp\left(-\frac{{\rm const} J_1}{(J_2-{J_2}_c)}\right)$
and a gap which is proportional to $\delta^2$.

\begin{figure}[bt!]
\centering
\epsfig{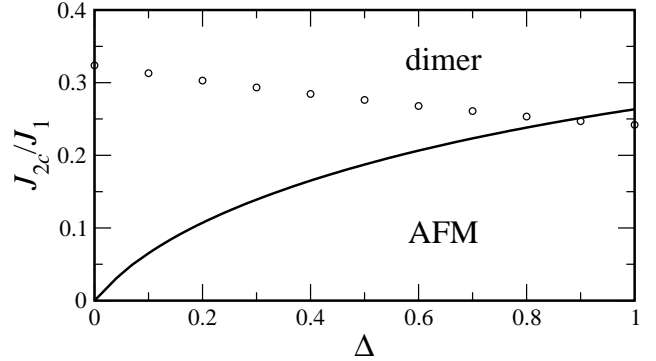}
\caption[]
{Phase diagram for the $J_1-J_2$ $XXZ$ chain as a function of
the anisotropy $\Delta$: the solid line is the result of
the mean-field approximation, the circles are
results of exact diagonalization \cite{somma2001}.
}
\label{phasediag}
\end{figure}

For an analysis of the phase diagram
we compare the ground state energies of all phases.
In the regime of small frustration,
the ``{\em AFM+stag.}'' solution
has the lowest energy and at some point
$\alpha_c={J_2}_c / J_1$ crosses with
the solution for the dimer state (see Fig.~\ref{gs-energy}
for $\Delta=1$). This can be identified with
a first-order phase transition. Note that at $\Delta = 1$ the ``{\em AFM+uni.}''
and the dimer solutions cross very close to  an
early estimate for the critical point \cite{haldane1982} $J_2=J_1/6$.
Taking into account the ``{\em AFM+stag.}'' solution shifts this crossing
point, thus yielding very good agreement with numerical
results \cite{eggert1996} ${J_2}_c\approx 0.242J_1$ for $\Delta=1$.
Fig.~\ref{phasediag} shows this crossing point as a
function of the anisotropy  $\Delta$. Although
the critical value of $J_2$ is very close to
numerical results for $\Delta\approx 1$,
the position of the crossing decreases with
decreasing $\Delta$ whereas the
exact diagonalization data \cite{nomura1993,somma2001} exhibit
an increase of ${J_2}_c$.

\begin{figure}[bt!]
\centering
\epsfig{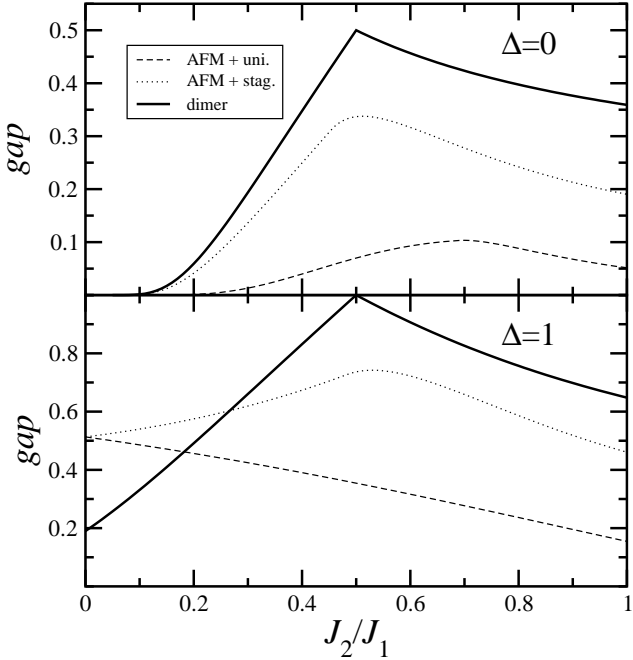}
\caption[]
{Excitation gap in the mean-field Hamiltonian
as a function of $J_2/J_1$
for the $J_1-J_2$ $XY$ chain (upper panel)
and the Heisenberg chain (lower panel).}
\label{gap}
\end{figure}

The excitation gaps of the different mean-field solutions
are depicted in Fig.~\ref{gap}.
The ``{\em uniform}'' solution is always gapless and is not displayed.
The ``{\em dimer}'' and ``{\em AFM+stag.}''solution have a maximum of the
gap near the Majumdar-Ghosh point which is consistent with numerical
results \cite{white1996}.
The position of the minimum in the excitation spectrum of
the ``{\em dimer}'' solution jumps from $k={\pm\pi}/{2}$
to $k=0$ at the
Majumdar-Ghosh point. The minimum of the ``{\em AFM+stag.}'' solution shifts
continuously for $J_2 > J_1/2$.

\begin{figure}[t!]
\centering
\epsfig{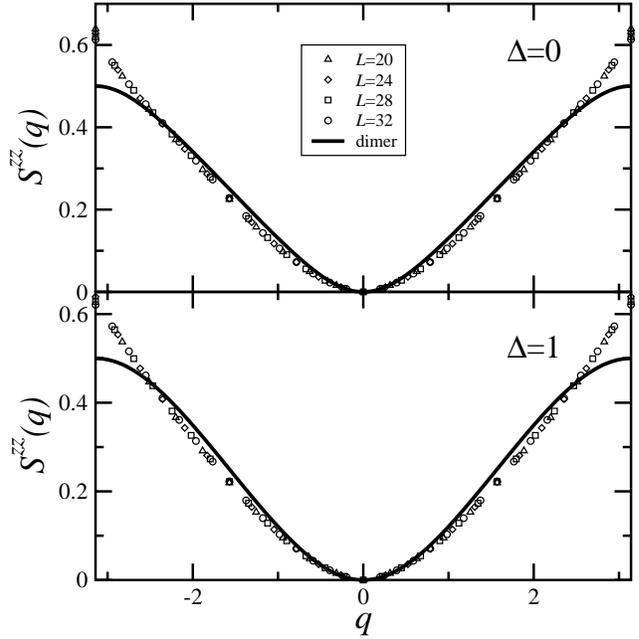}
\caption[]
{The structure factor $S^{zz}(q)$ as a function of $q$
for the $J_1-J_2$ $XY$ chain (upper panel) and Heisenberg chain (lower panel)
at $J_2 =5 J_1/11$.
The line is the result of the dimer mean-field solution;
symbols are exact diagonalization data for different system sizes
from 20 to 32 sites.
}
\label{figStruc}
\end{figure}

The $zz$-structure factor
$S^{zz}(q){=}\sum_{l=1}^L\exp(iql)\langle s_n^z s_{n+l}^z\rangle$
can be expressed through the four-fermion correlation function.
For the dimer state we get the following result in the mean-field
approximation:
\begin{eqnarray}
S^{zz}(q)=\frac{1}{4}
-\frac{1}{4\pi}\int_0^{\pi}dk \,
\frac{e_ke_{k+q}+f_k f_{k+q}}{\sqrt{(e_{k}^2+f_{k}^2)(e_{k+q}^2+f_{k+q}^2)}}.
\label{struct}
\end{eqnarray}
In the limit of the Majumdar-Ghosh model
the exact result \cite{bursill1995} $S^{zz}(q)=\frac{1}{4}(1-\cos(q))$
is recovered.
The result (\ref{struct}) for the $zz$ static structure factor
at $J_2/J_1=5/11$ is shown in Fig.~\ref{figStruc} as a full line
in comparison with numerical results (open symbols). Both approaches
yield a broad maximum at the boundary of the Brillouin zone $q= \pm \pi$,
signifying antiferromagnetic correlations with a short correlation
length \cite{white1996,bursill1995}.
Small deviations are observed at $q= \pm \pi$, where the numerical
data exhibits also the biggest finite-size effects.

\section{Conclusions}
\label{secConc}

To summarize we have applied the Jordan-Wigner transformation to
the spin-1/2 $J_1-J_2$ $XXZ$ chain.
We have performed a complete analysis of the possible mean-field states
and found that a dimerized state has lowest energy for $J_2$ larger
than some critical value. The ground-state energies are in good agreement
with exact diagonalization data and we even recover the exact result
at the Majumdar-Ghosh point \cite{MaGo}.

The location of the critical point is obtained from
the crossing between the energies of
the antiferromagnetically ordered state iii)
and the dimer state iv). This yields good agreement with numerical
results for $\Delta \approx 1$, but does not
follow the anisotropy dependence for small $\Delta$ correctly.
This discrepancy may be related to the following facts. Firstly,
spin-spin correlations should decay as a power-law below
${J_2}_c$ \cite{haldane1982}, while the mean-field theory treats this phase
as antiferromagnetically ordered. Secondly, in the mean-field scenario
the transition is generically of first order, not of the established infinite
order. Just at $\Delta = 0$, we find ${J_2}_c = \Jtd = 0$
and recover the infinite order phase transition to the dimer state,
see (\ref{asymptotic}).

Furthermore, we have calculated
the $zz$ static structure factor for the ``{\em dimer}''
mean-field approximation. The result coincides with the exact one for
the Majumdar-Ghosh point and satisfactory agreement is found with
numerical results in the dimerized phase below the Majumdar-Ghosh point.

It should be mentioned that there are problems with the present approach
for large $J_2$.
In particular, the mean-field dimer
state does not reveal any incommensurability, as signaled e.g.\ by
a shift of the maxima of the static structure factor beyond the
Majumdar-Ghosh point \cite{bursill1995}. This is an important difference
with Ref.\ \cite{sun2002} which obtained incommensurability from
a mean-field approach. However, the solution of Ref.\ \cite{sun2002}
does not correspond to the ground state. Remarkably, the
absence of incommensurability
is a special property of the dimerized mean-field solution which leads
to a cancellation of the next-nearest neighbor hopping processes in the
fermionic picture.
Consequently, alternative approaches are needed to describe
the region $J_2 > J_1/2$.

\begin{acknowledgement}

T.V.\ would like to thank the Physics Department of the TU Braunschweig for
hospitality during the course of this work, and the Deutsche
Forschungsgemeinschaft for financial support of this visit.
We would like to thank H.\ B\"uttner, O.\ Derzhko, Yu.\ Gaididei
for helpful discussions and
G.\ Bouzerar for communicating his results of a real-space
Jordan-Wigner mean-field calculation.
Some of the numerical computations have been performed on a COMPAQ ES45
(CFGAUSS) at the Rechenzentrum of the TU Braunschweig.

\end{acknowledgement}


\end{document}